# On the possible mechanism of superconductivity in cuprates and pnictides

K.V. Mitsen[1] and O.M. Ivanenko

Lebedev Physical Institute RAS, 119991 Moscow, Russia

**The nature of the normal state and the mechanism of superconductivity in two families of high-temperature superconductors, cuprates and pnictides, remain a matter of intense discussions. According to band-structure calculations, confirmed by ARPES data, the electronic structure of pnictides differs considerably from that of cuprates. However, in spite of these differences, it looks like in both cases there exists some general and fairly "coarse" mechanism independent of fine details of the band structure and responsible for superconductive pairing in these materials. Here, we suggest a qualitative model describing the ground state and the mechanism of superconductive pairing in cuprates and pnictides that, in our opinion, can explain many of their unusual properties.**

It is commonly agreed that, in high-temperature superconductors (HTSCs) based on iron pnictides, only Fe states appear at the Fermi level and it is these states that are responsible for transport, galvanomagnetic, and superconducting properties of these compounds. Meanwhile, the As (or Se) states, according to calculations [1], are ~2 eV below the Fermi level. This means that strong electron correlations at the cation, which determine the band structure of cuprates, are absent in pnictides.

At the same time, certain traits typical of both cuprates and pnictides are readily seen; these are the presence of quasi-two-dimensional layers of 3d metal cations (Cu and Fe, respectively) and ligand anions (O and As), as well as the existence of a band gap $\Delta_{ib}$ ~ 2 eV between the occupied ligand states and unoccupied cation band in the electronic spectra of the two families of compounds. With both families being high-temperature superconductors, it seems reasonable to relate this fact to features they have in common.

---

[1] -mail address: mitsen@sci.lebedev.ru

The above-mentioned feature of the crystal composition of cuprates and pnictides entails, in particular, a large contribution of the bulk Madelung energy $|E_M|$ to the electronic structure of anion-cation planes. We believe this circumstance to be of decisive importance both for the peculiarities of the ground (normal) state of these materials that appear upon doping and the unusual mechanism of Coulomb pairing arising in this ground state.

The chemical formulas of cuprates and pnictides suggest that the main contribution to the Madelung energy in these compounds is related to the interaction of positively charged cations ($Cu^{2+}$, $Fe^{2+}$) and negatively charged anions ($O^{2-}$, $As^{3-}$). At the same time, in cuprates as well as pnictides, the anion bands are occupied and the transfer of an electron from an anion to a cation requires an energy $\Delta_{ib}$ (Figs. 1a, 1d). In the case of cuprates (Fig. 1a), it is thought that $\Delta_{ib}$ is related to the Coulomb correlation of electrons at Cu ions; in the case of pnictides (Fig. 1d), $\Delta_{ib}$ is the band gap. However, in both cases it is possible to control the value of $\Delta_{ib}$ changing $|E_M|$.

In the context of the ionic model, the energy $\Delta$, related to the electron transfer from an anion to a cation is determined by the sum of three quantities [3]:

$$\Delta \approx |\Delta E_M| - I_d + A_p, \qquad (1)$$

Here, $|\Delta E_M|$ is the difference in the Madelung energies $|E_{M2}|$ and $|E_{M1}|$ of the two configurations where the anion and cation charge states differ by ±1, which correspond to the transfer of an electron from an anion to a cation; $I_d$ is the ionization potential of the cation; and $A_p$ is the electronegativity of the anion.

Now, let $\Delta_{ib}$ to be somehow reduced due to a reduction in $|E_{M2}|$. This is the case, for example, if a localized positive (negative) charge of appropriate magnitude is put in the vicinity of each cation (anion). We believe that in HTSCs this occurs through doping [4]. It is implied that charges introduced upon doping are localized in the immediate neighborhood of dopant ions, either at anions or cations within the plane (for $La_{2-x}Sr_xCuO_4$, $Nd_{2-x}Ce_xCuO_4$, $Ba(Fe_{1-x}Co_x)_2As_2$, etc.) or outside the plane at the nearest adjacent ions (for $YBa_2Cu_3O_7$,

$Bi_2Sr_2Ca_nCu_{n+1}O_y$, etc.). As far as the charge screening, effective at distances exceeding interatomic spacing, may be considered negligible at smaller distances, these charges are seen virtually unscreened by nearest-neighbor ions. Estimates indicate [5] that a local decrease in $|E_{M2}|$ for a given ion caused by one neighboring charge introduced by doping is about 1.5-2 eV.

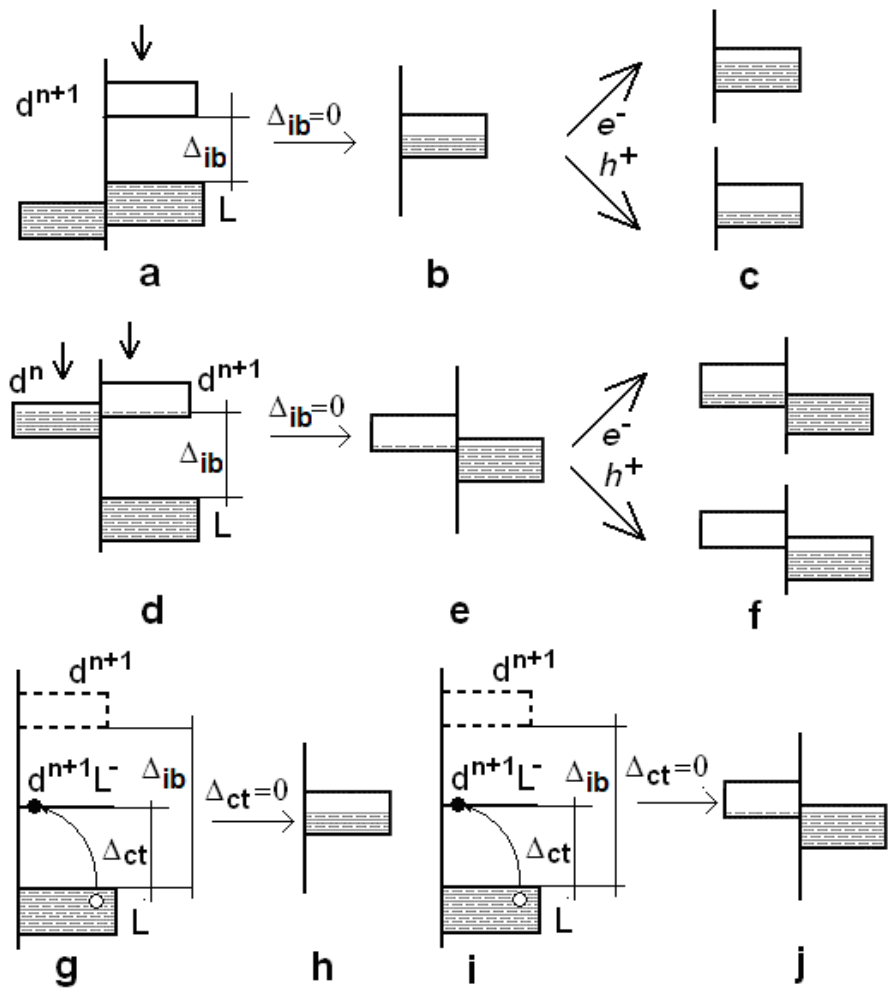

Figure 1. Modification of electron structure of cuprates and pnictides under doping; a, d – electron structures of undoped cuprates and pnictides, correspondingly: b, e – modification of electron spectra "a" and "b" if $\Delta_{ib}$ is reduced somehow to zero; c,f - modification of electron spectra "a" and "b" if $\Delta_{ib}$ is reduced to zero by electron or hole doping; g, i Minimal energy of interband excitation in cuprates and pnictides is the energy of exciton-like excitation $\Delta_{ct}$, that corresponds to the transfer of an electron from anion to neighboring cation with the formation of localized hole.

If $|E_{M2}|$ decreases to such an extent that $\Delta_{ib}$ be reduced to zero, new bands, formed by hybridized states of d and L bands, appear at the Fermi level (Figs. 1b, 1e). The resulting new conduction band can be either half-occupied (as in cuprates (Fig. 1b), with odd number of electrons per unit cell) or partially occupied (as in pnictides (Fig. 1e), with even number of electrons per unit cell under the condition of band overlap). Thus, the material will be a metal in both cases. For hole (electron) doping, this metal will exhibit electron (hole) conduction in the first case (Fig. 1c) and hole (electron) conduction in the second case (Fig. 1f). This is in full agreement with the types of conduction observed in overdoped phases of cuprates and pnictides upon hole and electron doping

Next, for the sake of common description of cuprates and pnictides, we disregard the possibility of transport via Fe cations in the latter case; i.e., we assume that the upper band is empty in both cases. This restriction does not affect the essence of our treatment.

In both cases, an excitation with the energy $\Delta_{ib}$ corresponds to the transfer of an electron from a ligand anion to a cation with the formation of a band hole. However, it is another, exciton-like excitation $3d^{n+1}(L^-)$, which has the lowest energy $\Delta_{ct} < \Delta_{ib}$. This excitation corresponds to the transfer of an electron from a ligand anion (O, As) to the nearest 3d cation (Cu, Fe) with the formation of an $L^-$ hole localized at the neighboring ligands (Figs. 1g, 1i). We assume that interaction of this hole with the neighboring cation is unscreened. Thus, as $|E_{M2}|$ and $\Delta_{ib}$ are gradually reduced, one expects that a state with $\Delta_{ct} = 0$ be attained first.

Let a state with $\Delta_{ct} = 0$ for the entire anion-cation plane be attained. We will assume that the *p-d* hybridization prevents a possible transition into an excitonic insulator state [6]. Then, we have a system with one half-occupied band for cuprates (Fig. 1h) and two overlapping bands (one electron-type and another hole-type) for pnictides (Fig. 1j). Extended electron states in these bands are formed from $3d^{n+1}(L^-)$ and $(3d^n)L$ states. In such a system, transitions of electrons from the anions can take place only to the nearest-neighbor cations and back, but not to the next-neighboring ions. I.e., there still exists a gap for the charge transport between the neighboring cells and, consequently, incoherent charge transport within the plane along anion-cation bonds cannot occur. At the same time, this system possesses an constant-energy Fermi surface (FS), similarly to metals, insofar as nothing prevents electrons from "flowing" in the momentum space so that energy is changed but the number of electrons in a unit cell remains constant. For cuprates, there exists one FS (Fig. 1h); for pnictides, there are two FSs (an electron and a hole ones) with equal concentrations of the carriers (apart from those introduced by doping) (Fig. 1i). Despite the fact that incoherent transport in such a system cannot take place, the existence of a FS leads to the possibility of coherent transport, when the entire electron system moves as a whole (a condensate).

Let us now examine a possible mechanism whereby such a coherent (superconducting) state may be established. Consider a system where $\Delta_{ct} = 0$ and electron states at the Fermi level are formed from $3d^{n+1}(L^-)$ and $(3d^n)L$ states. This means that whenever an electron appears on a cation, it leaves behind a hole on a neighboring anion. It is easily understood that, in this case, two electrons on neighboring cations have a lower energy as they experience attraction to two holes simultaneously. Thus, each pair of cations in such a system may be considered as a two-atom negative-U center (NUC) [4]. Owing to virtual transitions of electron pairs to these NUCs, states ($k\uparrow,-k\downarrow$) in the vicinity of the FS are pairwise coupled, which leads to superconducting pairing in the system.

In the normal state of the system under consideration, each electron has to be localized in its cell. At the same time, the two-particle hybridization results in redistribution of a part of electron density to NUC-localized states, with holes being left in anion orbitals [7]. An overlap between the wave functions of holes belonging to different NUCs leads to the formation of extended hole states, which provides for the hole transport via the anion sublattice of the crystal in the normal state. While in pnictides there additionally exists transport via the cation sublattice, in cuprates the above mechanism is the only one responsible for the charge transport in the normal state.